\documentclass[12pt, a4paper,reqno]{amsart}
    \usepackage{amsmath,amssymb,amstext} 
          \newcommand{\ind}{\perp\!\!\!\!\perp} 
    \usepackage[a4paper, margin=1in]{geometry} % Adjust margins as per journal requirements
    \usepackage{setspace} % For line spacing control
    \usepackage{booktabs, array}
    
    \usepackage{natbib} % If the journal requires specific citation styles

    \usepackage{hyperref} 
    
    \usepackage{rotating,multicol,multirow,paralist}
    
    \usepackage{tikz}
    \usetikzlibrary{arrows.meta}
    \usepackage{enumitem}

    \usepackage{algorithm}
    \usepackage{algorithmic}
    \usepackage{caption} % Include the caption package
\begin{document}
\title[]{Doubly robust Methods for Recurrent Event Outcomes: Causal Effects of Blood Pressure Medications on Acute Kidney Injuries}
\maketitle
\author{ 
\begin{minipage}{0.95\textwidth}
\centering \small
Wenling Zhang$^{1}$, Cecilia Cotton$^{2}$ and Lan Wen$^{2}$
\\ [2em]
$^{1}$ Department of Epidemiology, Harvard T.H. Chan School of Public Health, Boston, MA, USA\\

$^{2}$ Department of Statistics and Actuarial Science, University of Waterloo, Waterloo, Ontario, Canada

\end{minipage} 
}

\section*{Abstract}

{Evaluating the average causal effects of treatment strategies on recurrent event outcomes, such as heart attacks or renal failure, is important in clinical and medical research. However, the analysis becomes increasingly complex as multiple interacting factors are considered within a longitudinal setting. In this paper, we use advanced methodologies to estimate the average causal effects of standard versus intensive blood pressure–lowering therapies on acute kidney injury recurrences. We address time-varying treatment and confounding, and model misspecification during the identification and estimation processes for the effect estimands. We analyze the Systolic Blood Pressure Intervention Trial data set using our proposed method, accounting for medication adherence and the (semi-)competing risk of death observed in the data.}

\allowdisplaybreaks\raggedbottom
\section{Introduction}

Hypertension is considered a major risk factor for multiple recurrent diseases, including heart disease, stroke, and kidney failure, as it contributes to arterial damage, reduced blood flow, and organ dysfunction over time \cite{chobanian2003seventh}. Nevertheless, in blood pressure management, challenges emerge in balancing the benefits of maintaining stable blood pressure with the risks of potential renal damage. 

The Systolic Blood Pressure Intervention Trial (SPRINT) \cite{wright2016randomized} is an open-label clinical trial designed to evaluate the effects of intensive versus standard blood pressure medication therapy in hypertensive adults 50 years of age or older.
{The SPRINT research group} found that patients \textit{assigned} to the intensive treatment group experienced higher rates of acute kidney injury (AKI) episodes compared to their standard treatment counterparts, after a median follow-up of 3.26 years \cite{wright2016randomized}. However, the intensive group also saw a reduction in all-cause mortality rates compared to the standard group \cite{cheung2017effects, bress2017potential}. 
While antihypertensive medications aim to protect cardiovascular health, research indicates that their effects on kidney function are complex and can vary significantly across patients \cite{hamrahian2017hypertension}. Despite recent guidelines, the optimal blood pressure target for treating hypertension patients remains controversial \cite{arguedas2020blood}. 

%{Existing Methods}
Previous studies estimating the average causal effect (ACE) for recurrent events have predominantly focused on a point exposure setting \cite{Amorim2015,gao2016causal,Su2020, su2022causal}. However, not all treatment strategies can be analysed as point treatments, particularly when considering the compliance with sustained treatment measures. In the case of prescribed medications, the actual implementation of treatment may not be consistent due to variations in patients' medication adherence over time. In studies where non-adherence to treatment medication is a potential issue, researchers may choose to conduct an intent-to-treat (ITT) analysis and/or a per-protocol (PP) analysis \cite{hernan2017per, porta2007discordance}. Motivated by the SPRINT, our interest lies in examining the ACE of standard versus intensive medication treatments for blood pressure control on the recurrence of acute kidney injury (AKI) episodes using robust methodology for both ITT and PP analyses. Understanding this research question is crucial for clinicians in making informed decisions to reduce the frequency of AKIs in hypertensive populations and ultimately improve their quality of life.

{
We adopt the counterfactual framework to define our causal effects of interest \cite{hernanbook, Rubin1974, holland1986, Pearl2009} and build on the longitudinal g-formula \cite{robins1986new}. 
We extend the longitudinal targeted maximum likelihood estimation (TMLE; \cite{van2006targeted, van2011targeted, lendle2017ltmle, schomaker2019using, hoffman2024studying}) to handle recurrent event outcomes on a discrete time grid, and provide a transparent implementation to facilitate its application in real-world studies. In particular, we (i) formalize the target causal estimands and the corresponding longitudinal data structure for recurrent event outcomes under time-varying treatment, (ii) provide practical guidance on constructing longitudinal outcomes from recurrent events and interpreting the resulting estimands, and (iii) illustrate these methods through a comprehensive analysis of the SPRINT data. Building on the theory of efficient influence functions \cite{tsiatis2006semiparametric,levy_tuteif_2019, hines2022demystifying}, we also describe the TMLE implementation, highlighting its double-robustness properties and its ability to incorporate machine-learning approaches for nuisance-function estimation.
}
\section{Observed Data Structure and Causal Identification}
\label{sec:identification}

Consider a longitudinal study with $N$ subjects sampled from a target population of interest. Suppose the time intervals are defined fine enough such that the chances of experiencing more than one recurrent event during a single interval is very small. Throughout, let $k$ index time intervals, where $k = 0$ denotes baseline and $k = K+1$ denotes the end of follow-up. Each interval $k$ begins at and includes time point $k$, and ends just before time point $k+1$. Let $L^*_0$ denote a vector of baseline variables measured at the beginning of the study period ($k = 0$) that may influence treatment and the outcome of interest throughout the study. For each time interval $k \in \{1,...,K\}$, let $L^*_k$ represent a vector of time-varying covariates measured at $k$, which may affect treatment and the outcome during that specific interval. 
Note that $L^*_0$ and $L^*_k$ may consist of different sets of variables. 

{Motivated by the SPRINT setting, let $Z \in \{0,1\}$ denote baseline randomized treatment assignment in a trial, where $Z=1$ indicates assignment to the treatment and $Z=0$ indicates assignment to the control. For $k \in \{1,\ldots,K\}$, let $A_k \in \{0,1\}$ denote adherence to the assigned strategy during interval $k$, with $A_k=1$ indicating adherence to the regimen specified by $Z$ throughout that interval.}

Let $Y_k$ denote an observed binary variable indicating whether an event occurred during time interval $k$ (i.e., $Y_k \in \{0,1\}$), and let $R_{K+1}$ denote the corresponding observed cumulative sum of events. Specifically, $R_{K+1} = \sum_{j=1}^{K+1} Y_k$ represents the cumulative count of recurrent events over the study period. By definition, $Y_0 = R_0 = 0$. 
{For simplicity, we assume no loss to follow-up in the main text, with censoring in more general scenarios discussed in Appendix A.1--A.4.}
\sloppy

For notational convenience, let the vector $L_{k}$ denote an individual's joint time-varying covariates, which is composed of the binary outcome variable $Y_{k}$ and the vector of covariates $L^*_{k}$ in interval $k$, i.e., $L_{k} = (Y_{k}, L^*_{k}),\ k \in \{1,...,K\}$. For completeness, define $L_0= L^*_0$ and $L_{K+1}= Y_{K+1}$. Following this notation, the assumed observed data structure is given by:
{$O = (L_0, Z, L_1 , A_1 ,..., L_K , A_K, Y_{K+1})\sim P,$}
\noindent where $P$ is the true population distribution of the observed data. Table \ref{tab:sec2_notation} summarizes the key notation introduced in this section. For notational simplicity, we assume all variables are discrete; if continuous variables are present, summations would be replaced with integrals as appropriate. 

{We define an intervention path $\bar{a}_K = (a_1, \ldots, a_K)$ as a deterministic sequence of adherence indicators under the assigned regimen $Z=z$, ${a}_k \in \{0,1\}$ for $k \in \{1,...,K\}$. In addition, denote the interval-specific counterfactual risk under assignment $z$ and adherence history $\bar{a}_{k-1}$ through interval $k-1$ by $\psi_{k}^{z,\bar{a}_{k-1}} \equiv \mathbb E \big(Y_k^{z,\bar{a}_{k-1}}\big),$ where $\bar{a}_{k-1}=(a_1,\ldots,a_{k-1})$ is a user-specified adherence regime. Corresponding causal contrasts may compare different values of $z$ and/or different adherence regimes $\bar{a}_{k-1}$.} % with the per-protocol contrast obtained as the special case $\bar{a}_{k-1}=1$.} 
 
 To identify our causal estimand, we invoke the following causal assumptions $\forall k$ \cite{hernanbook, robins1986new}:

\begin{enumerate}
     \item Conditional Exchangeability: $\bar{Y}_{K+1}^{\bar{a}}\ind Z\mid L_0$ and 
    $\{Y_k^{z,\bar{a}_{k-1}},..., Y_{K+1}^{z,\bar{a}_{K}}\} \ind  A_{k-1} \mid \bar{L}_{k-1}, {Z=z}, \bar{A}_{k-2}=\bar{a}_{k-2}.$
    \item Positivity: $\Pr( A_{k-1}=a_{k-1}\mid \bar{L}_{k-1}=\bar{l}_{k-1}, {Z=z}, \bar{A}_{k-2}=\bar{a}_{k-2})>0$ and {$P(Z=z\mid L_0)>0$} for all observed covariate values with positive density under the intervention $(z,\bar{a}_{k-1})$.
    \item Consistency: {If $Z = z$} and $\bar{A}_{k-1}=\bar{a}_{k-1}$, then $\bar{Y}_k = \bar{Y}_k^{z,\bar{a}_{k-1}}$.
\end{enumerate}

{
In the SPRINT trial, exchangeability, positivity and consistency for \(Z\) is ensured by baseline randomization. Exchangeability for adherence requires that, conditional on the observed history \(\bar L_{k-1}\), baseline assignment \(Z=z\), and prior adherence \(\bar A_{k-2}=\bar a_{k-2}\), there are no unmeasured common causes of adherence at interval \(k-1\) and future outcomes; while this assumption is not empirically testable, its plausibility depends on the richness of the measured covariate history. 
Positivity for adherence requires that each adherence level specified by the intervention has a positive probability of occurring within strata defined by the observed history, which can be assessed by examining the empirical distribution of adherence across covariate strata. Finally, consistency for adherence requires that adherence indicators \(A_k\) are well-defined under interventions; this assumption is not empirically testable and instead relies on the clarity of the causal question of interest \cite{hernan2016does,young2024story}. }

These assumptions allow us to represent the estimand in an iteratively conditional expectation (ICE) form given by:
\begin{equation}
\label{eqn:tmle2_ice_form}
\begin{aligned}
\psi_{k}^{z,\bar{a}_{k-1}}
&\equiv \mathbb E \bigl(Y_k^{z,\bar{a}_{k-1}}\bigr) \\
&=
\mathbb E\Bigl[
\mathbb E\Bigl(
\cdots
\mathbb E\bigl(
Y_k
\mid
\bar L_{k-1}, {Z=z}, \bar A_{k-1}=\bar a_{k-1}
\bigr)
\Bigr]
\cdots
\mid
L_0, {Z=z}
\Bigr)
\Bigr].
\end{aligned}
\end{equation}
which, by linearity, identifies
\begin{equation}
\label{eqn:tmle2_ice_form_R}
\begin{aligned}
\mathbb E \bigl(R_{K+1}^{z,\bar a_K}\bigr)
&= \sum_{j=1}^{K+1} \psi_j^{z,\bar a_{j-1}} \\
&= \sum_{j=1}^{K+1}
\mathbb E\Bigl[
\mathbb E\Bigl(
\cdots
\mathbb E\Bigl[
\mathbb E\bigl(
Y_j
\mid
\bar L_{j-1}, {Z=z}, \bar A_{j-1}=\bar a_{j-1}
\bigr)
\mid
\bar L_{j-2}, {Z=z}, \bar A_{j-2}=\bar a_{j-2}
\Bigr]\\
& \qquad\qquad \qquad
\cdots
\mid
L_0, {Z=z}
\Bigr)
\Bigr].
\end{aligned}
\end{equation}

Detailed identification proofs are provided in Appendix A.1, where we adopt a more general framework that accommodates both trial and observational settings; the resulting formulation is equivalent and the methodology remains unchanged.
 
\section{Methodology}
\label{sec:tmle2_method}

In this section, we provide the necessary steps to construct doubly robust estimators based on the influence function of \eqref{eqn:tmle2_ice_form}. Details of the nuisance models and the estimation processes for a suite of other comparative estimators are provided in Appendix A.3. To begin, we first define the nuisance models for the propensity score and the outcome processes as follows.

\subsection{Nuisance Models} 
\label{sec:tmle2_method_nuisance}

{To aid in estimation, let $\bar\pi^{z,\bar a_k}=\pi^{z}\prod_{j=1}^k \pi^{z,a_j}$ denote the joint probability of baseline treatment assignment and subsequent adherence through interval $k$, where $\pi^{z}=\Pr(Z=z\mid L_0)$, and $\pi^{z,a_j}=\Pr(A_j=a_j\mid \bar L_{j-1},Z=z,\bar A_{j-1}=\bar a_{j-1})$.}
Furthermore, define ${Q}_{k,k}^{z,\bar{a}_{k-1}} \equiv Y_k$, and iteratively backwards in time, let
{${Q}_{k,h}^{z,\bar{a}_h}\equiv \mathbb E(Q_{k, h+1}^{z,\bar{a}_{h+1}}\mid Z=z, \bar{A}_{h}=\bar{a}_{h},\bar{L}_{h})$}
for $h <k$, as given by the identifying formula \eqref{eqn:tmle2_ice_form}.

\subsection{Efficient Influence Function}
In robust statistics, an influence function measures how sensitive an estimator is in responding to the changes to each individual data point. In particular, an efficient influence function (EIF) is the influence function that achieves a semiparametric efficiency bound \cite{tsiatis2006semiparametric, levy_tuteif_2019, hines2022demystifying,hampel1974influence}. 
Based on the EIF, we are often able to construct estimators with desirable properties, such as double robustness. 

{By taking the pathwise derivative of $\psi_{k}^{z,\bar{a}_{k-1}}$ along a parametric submodel of the observed data distribution and summing up over $k = 1,...,K+1$, the EIF for $\mathbb E (R_{K+1}^{z,\bar{a}_{K}})$ identified by \eqref{eqn:tmle2_ice_form_R} is given by:}
\begin{align}
\xi_{K+1}^{*\,z,\bar a_K}
&=
\sum_{k=1}^{K+1}\sum_{j=1}^{k}
\Biggl[
\frac{\mathbb{I}\!\left(Z=z,\ \bar A_{j-1}=\bar a_{j-1}\right)}
{\bar{\pi}^{z,\bar a_{j-1}}}
\bigl\{Y_j-Q_{j,j-1}^{z,\bar a_{j-1}}\bigr\}
    \label{eqn:tmle2_eif}
\\
&\qquad\qquad \qquad
+\sum_{m=2}^{j}
\frac{\mathbb{I}\!\left(Z=z,\ \bar A_{m-2}=\bar a_{m-2}\right)}
{\bar{\pi}^{z,\bar a_{m-2}}}
\bigl(Q_{j,m-1}^{z,\bar a_{m-1}}-Q_{j,m-2}^{z,\bar a_{m-2}}\bigr)
+ Q_{j,0}^{z,a_0}
-\psi_{j}^{z,\bar a_{j-1}}
\Biggr].\nonumber
\end{align}
Derivation details are provided in Appendix A.2.
\sloppy
Based on the this EIF, we construct doubly robust estimators that remain consistent as long as at least one of the (a) propensity score or (b) outcome model is correctly specified. 

\subsection{{Targeted maximum likelihood (TML) estimator for interim outcome $\psi_{k}^{z,\bar{a}_{k-1}}$}}
\label{sec:tmle2_method_tml}
As an EIF-based solution for estimating the expected potential outcome mean \cite{van2011targeted}, a TML estimator includes an additional update process for the ease of incorporating machine learning algorithms. In TMLE, we update the outcome regression models to account for the residual bias due to potential model misspecification, for example, by fitting a propensity score-weighted regression to estimate a fluctuation parameter. The detailed estimation steps are given by Algorithm \ref{alg:TML_procedure}. Due the complex nature of the estimator, detailed estimating process will be given in $k=2$ for conciseness. {General procedures for an arbitrary length of the study period are provided in Appendix A.3.}

\begin{table}[htb!]
\centering
\begin{minipage}{\linewidth}

\captionsetup{labelformat=empty} 
\hrulefill
\caption{\textbf{Algorithm 1: Targeted maximum likelihood estimation (TMLE)}}
\captionsetup{labelformat=default}
\label{alg:TML_procedure}
\begin{algorithmic}[1]
    \STATE Obtain $\bar{\hat{\pi}}^{z,{a}_1}$, an estimate of $\bar{{\pi}}^{z,a_1}$ as described in Section \ref{sec:tmle2_method_nuisance}.
    
    \STATE Select individuals that are uncensored until time interval $2$ and who followed path $(z, {a}_{1})$:
    \begin{enumerate}
        \item[i.] Regress $Y_2$ on past measured variables $\bar{L}_{1}$ using the selected data, denote the model as $F_a$.
        \item[ii.] Obtain an initial estimate of ${Q}_{2,1}$ by fitting $F_a$, denoted as $\hat{Q}_{2,1}$.
        \item[iii.] Regress $Y_2$ on $1$ with an offset (fixed intercept) of the initial estimate logit $\hat{Q}_{2,1}$, where each observation in the score function is weighted by $[\bar{\hat{\pi}}^{z,a_{1}}]^{-1}$; denote the new model as $F_a^*$.
    \end{enumerate}
    
    \STATE Select individuals with $A_1=a_1$ until time interval $1$ with  $Z=z$:
    \begin{enumerate}
        \item[i.] Obtain the fitted values, denoted as $Q^*$, using model $F_a$ with the newly selected data.
        \item[ii.] Obtain the updated estimate $\hat{Q}^{*}_{2,1}$ by fitting $F_a^*$ using the new offset logit $Q^*$ with the selected data.
    \end{enumerate}
    
    \STATE In the same individuals:
    \begin{enumerate}
        \item[i.] Regress $\hat{Q}^{*}_{2,1}$ on ${L}_{0}$, denote the model as $F_b$.
        \item[ii.] Obtain the fitted values with model $F_b$, denoted as $\hat{Q}_{2,0}$.
        \item[iii.] Regress $\hat{Q}^{ *}_{2,1}$ on $1$ with the offset of logit $\hat{Q}_{2,0}$ where each observation in the score function is weighted by $[{\hat{\pi}}^{z}]^{-1}$; denote the new model as $F_b^*$.
    \end{enumerate}
    
    \STATE For all individuals (unstratified),
    \begin{enumerate}
        \item[i.] Obtain a new offset term using model $F_b$, denoted as $Q^{**}$.
        \item[ii.] Obtain the updated estimate $\hat{Q}^{  *}_{2,0}$ by fitting $F_b^*$ using new offset logit $Q^{**}$.
    \end{enumerate}
    
    \STATE Calculate $\hat \psi_{2}^{z, a_{1}}=\mathbb{P}_n(\hat{Q}^{*}_{2,0})$ using all observations in the study.
\end{algorithmic}
\hrulefill

\end{minipage}
\end{table}
%(Table of Estimation Procedure 1)

\subsection{Estimation of causal contrast}

{After estimating $\psi_{k}^{z,\bar{a}_{k-1}}$ for $k=1,\ldots,K+1$, we can construct causal contrasts of interest by aggregating the corresponding estimates under prespecified intervention regimes. In general, for two regimes $(z^\dag,\bar a_{k-1}^\dag)$ and $(z^{\dagger \dagger},\bar a_{k-1}^{\dagger\dagger})$, let
$$\hat{\Psi}=\sum_{k=1}^{K+1}\hat{\psi}_{k}^{z^\dag,\bar a_{k-1}^\dag}-\sum_{k=1}^{K+1}\hat{\psi}_{k}^{z^{\dagger\dagger},\bar a_{k-1}^{\dagger\dagger}}.$$

{Appendix A.4 presents simulations illustrating the finite-sample performance of the longitudinal estimator.}

%\FloatBarrier
\section{Systolic Blood Pressure Intervention Trial (SPRINT)}
\label{sec:tmle2_apply}

{Using data from the SPRINT study, we estimate the causal effects of intensive versus standard blood pressure lowering therapy on recurrent AKI episodes. We consider several estimands that address different aspects of this question, including intention-to-treat (ITT), per-protocol (PP), and related effects defined by adherence and other time-varying factors as described below.}
We utilize machine learning algorithms in the doubly robust estimator described herein for effects estimation and comparison.

\subsection{Data Description}
\label{sec:tmle2_apply_setting}

The SPRINT cohort consists of 9,361 subjects who were selected based on their high systolic blood pressure ($\geq$ 130 mm Hg), elevated risk of cardiovascular disease, and absence of diabetes. At the beginning of the study, participants were randomly assigned to either a standard treatment group, with the objective of reducing systolic blood pressure to less than 140 mm Hg, or an intensive treatment group, targeting a systolic blood pressure of less than 120 mm Hg. 

Our analysis consists of 9,322 subjects with complete baseline variables. 
{Given the negligible missingness in baseline covariates (approximately 0.4\%), we assume that the data are missing completely at random, as any deviations are unlikely to meaningfully affect the estimates.} 

We define the first four years of the trial as our study period. During this period, no participants were right-censored due to loss to follow-up \cite{bellows2021estimating}. 
The observed all-cause mortality rate in our study was below 5\%, with cardiovascular-related deaths accounting for 32.3\% of all deaths. {Henceforth, we define $D_k$ as a participant's all-cause mortality indicator during time interval $k$.} 

We define other key variables as follows, with their corresponding notation in parentheses. In the SPRINT data analysis, baseline treatment assignment has two levels: the standard blood pressure lowering therapy ($Z=0$) and the intensive blood pressure lowering therapy ($Z = 1$). {The time-varying $A_k$ indexes participants' adherence level, where $A_k = 1$ indicates participants' full compliance to doctor-prescribed blood pressure lowering medications during interval $k$.}
{We define $Y_k$ as an indicator of whether a subject experiences an AKI episode during interval $k$, and $R_{K+1}$ as the total number of AKI episodes over the follow-up period.} The baseline confounders $L_0$ are defined as before. The subject's recent measurement of mean arterial pressure (MAP) over time $k$ ($L^*_k$) is considered as the time-varying confounder, estimated by the equation MAP $=$ DP + 1/3(SP $–$ DP) where SP and DP denote systolic and diastolic blood pressure, respectively, which is a standard practice in clinical studies \cite{meaney2000formula,demers2021physiology}. 

For ease of implementation, our study period of 1,460 days (four years) is divided into two time intervals, as detailed in Appendix A.5.2. Table \ref{tb:tmle2_tv_sprint2} shows the distributions of the key time-varying variables with respect to treatment assignment after data coarsening. Although some information is inevitably lost through this transformation process, the simplification enables a more straightforward analysis and interpretation while preserving the main characteristics of the data. The causal graphs illustrating the chronological order of key variables for both the ITT and PP analyses are provided by Figure 4 and Figure 5, respectively, in Appendix A.5.5. {
Since less than 0.24\% of subjects experienced more than one AKI episode within an interval (see Table 8 of Appendix A.5.2), within-interval multiple episodes are extremely rare. We therefore count only the first episode in each interval, which is expected to result in a negligible undercounting of the total number of AKI episodes over follow-up. Under this simplification, $R_{K+1}$ provides a reasonable approximation to the total number of AKI episodes over follow-up.} 

\subsection{Total effect}
\label{sec:sprint_te}

Since blood pressure medications are designed to reduce the risk of cardiovascular disease, treating death as a censoring event and defining an estimand under the hypothetical elimination of this censoring mechanism may overlook treatment effects mediated through fatal events. This approach can result in an analysis that does not accurately reflect real-world risk. Moreover, this hypothetical intervention is not well-defined (see Young \textit{et al.} \cite{young_compete_2020} for a detailed discussion).

Janvin \textit{et~al.} \cite{janvin2024causal} highlight the importance of considering death as a separate time-varying variable, which opens an additional causal path from treatments to recurrent outcomes. Motivated by this approach, we consider death as a semi-competing event to the primary outcome of acute kidney recurrences \cite{aalen2008survival}. In this semi-competing risks setting, our research interest focuses on the recurrent, non-terminal outcome of AKIs, whose counts are subject to the terminal event of death \cite{haneuse2016semi}. 

In the SPRINT application, {we aim to estimate the \textit{total effect} of treatment}, as it is likely to have broader public health implications compared with the controlled direct effect.

{To account for the presence of all-cause mortality, $D_k$ is not treated as a right-censoring mechanism eliminated by the intervention, but rather as a semi-competing risk that truncates the recurrent outcome process and removes individuals from the risk set for subsequent AKI episodes. Accordingly, $D_k$ is incorporated as a component of the time-varying covariates $L_k^\ast$, with a deterministic relationship with the interim outcome such that $D_k=1$ implies $Y_k=0$ thereafter. This ensures that the recurrent outcome remains well defined throughout follow-up.}

We apply the proposed TMLE algorithm specified in Section \ref{sec:tmle2_method} to the processed SPRINT dataset \cite{marschner2011glm2} {to estimate total ITT and total PP}. 
The standard errors used to construct confidence intervals are obtained via 1,000 nonparametric bootstrap samples. We calculate the estimates where the nuisance models are fitted using parametric generalized linear models (GLMs), extreme gradient boosting (XGB) algorithms \cite{chen2015xgboost}, and generalized additive models (GAMs) \cite{wood2017gam}, respectively.

{Under the assumptions described in Section \ref{sec:identification}, the total ITT effect of $\Psi_{ITT,{2}}(P) = \mathbb E_P(R_{2}^{z=1}- R_{2}^{z=0})$ based on individuals' assigned treatment group ($Z$) can be identified with $A_1=\emptyset$.}

Using TMLE with nuisance models estimated via parametric GLMs, the estimated total ITT effect during the first four years of the study was 0.019 (95\% confidence interval (CI): [0.011,0.026]). 
Hence, our results suggest that assigning individuals to an intensive blood pressure lowering target is expected to result in more AKI events than assigning them the standard target in the first four years of the SPRINT study.

In the total PP analysis, we are interested in the difference in the expected total acute kidney recurrences had all participants strictly adhered to an assigned intensive blood pressure lowering therapy, compared to if they had fully adhered to an assigned standard therapy. 
{Specifically, this causal contrast is given by $\Psi_{\text{PP},2}(P) =\mathbb E_P(Y_{2}^{z=1,{a}_{1}=1} - Y_{2}^{z=0,{a}_{1}=1})$.}

Following the settings in Section \ref{sec:tmle2_apply_setting}, the total PP effect of fully adhering to the assigned intensive therapy throughout the study period versus fully adhering to the assigned standard therapy on {the total number of AKI recurrences can be identified.}
{Using the proposed TMLE with parametric GLMs for nuisance models, the corresponding total PP cumulative contrast in the SPRINT analysis was estimated to be 0.017 (95\% CI of [0.007,0.027]).}
This implies that, following a PP analysis, SPRINT participants under full adherence to their assigned intensive blood pressure lowering therapy would expect marginally {higher probability of experiencing recurrent AKI events} within the first four years of the study compared to them fully adhering to the assigned standard therapy for blood pressure management. 

Note that the ITT and PP effect estimates are quite similar. One possible explanation is that full adherence to either intensive or standard treatment has a negligible impact on recurrent AKI events, as adherence to the assigned treatment was already high in the study population. This issue is explored further in the following section.

\subsection{Supplementary analyses}
\label{sec:sprint_supplementary}

\subsubsection{Population Intervention Effect of Adherence}
\label{sec:app_tmle2_apply_pie}
As a reasonable measure for treatment adherence effect on the recurrent outcome, we explore the population intervention effect (PIE) by comparing the average outcomes if the entire population adhered to their treatment assignment versus the observed average outcomes observed in the real-world population \cite{westreich2017patients, rogawski2022population} for each of the treatment groups. Appendix A.5.3 details the estimation results for the total PIEs in the intensive and standard arms, respectively. Based on these results, there is little evidence to suggest that full adherence to assigned intensive (or standard) blood pressure-lowering targets has a significant effect on recurrent AKIs, compared with subjects' observed adherence levels during the first four years of the study. These results align with findings in Section \ref{sec:sprint_te}.

\subsubsection{Average Controlled Direct Effect on Recurrent Outcomes}
In addition to the average total ITT effect estimated in Section \ref{sec:sprint_te}, we also examined the corresponding average controlled direct effect of intensive versus standard blood pressure lowering therapy on AKI recurrences. {See Appendix A.5.3 for further discussion of the effect definition.} Using parametric GLMs for the nuisance models, the estimated controlled direct effect was 0.018, with a 95\% CI of [0.010, 0.026]. Alternative estimates obtained using machine learning algorithms yielded similar effect sizes, as presented in Table 9 of Appendix A.5.4.

\subsubsection{Sensitivity Analysis of Time Discretization}
{We conducted a sensitivity analysis to assess whether the estimated causal effects were sensitive to the choice of temporal discretization. In addition to the primary bi-yearly coarsening, we re-analyzed the data under a finer annual coarsening scheme. The resulting estimates were highly consistent across discretization levels. In particular, the estimates under annual coarsening closely matched those from the primary bi-yearly analysis and were also comparable to those from the continuous-time inverse probability weighted ITT analysis of \cite{janvin2024causal}. These results suggest that the main causal conclusions are robust to the choice of temporal aggregation. Details of the sensitivity analysis, including the distributions of the coarsened variables, numerical results, and implementation, are provided in Appendix~A.5.6.}

\subsection{Comparative Results}
%Turning race and smoke status into binary variable.
For each of the estimands defined in the previous sections, Table \ref{table:sprint_est_compare} compares the effect estimates using the TML estimator with parametric GLMs and the TML estimator incorporating XGB algorithms \cite{chen2015xgboost}. In addition, Appendix A.5.4 shows the estimation results using GAMs to estimate the nuisance models, which closely resemble those shown in Table \ref{table:sprint_est_compare}.

Across both ITT and PP analyses, individuals assigned to an intensive blood pressure lowering target are expected to experience more AKI events than they would have under assigned standard treatment during the first four years of the SPRINT study. Notably, the total PP effect estimate is practically identical to those from the total ITT analysis. This is not surprising given that, when estimating the PIEs of adherence to prescribed blood pressure medications, as detailed in Section \ref{sec:app_tmle2_apply_pie}, the results indicate that the difference in AKI recurrences between full medication adherence and observed adherence is negligible, regardless of participants' assigned treatment arms.

\section{Discussion}
Motivated by the Systolic Blood Pressure Intervention Trial (SPRINT) dataset, we presented doubly robust estimators along with detailed estimation procedures tailored to our recurrent event outcome in complex settings involving time-varying treatments. Theoretical frameworks for these estimators were rigorously validated via simulation studies in Appendix A.4, demonstrating their applicability and robustness. 

We implemented the TMLE methodologies on the SPRINT data. 
The ITT analysis offered insights into the effectiveness of the assigned treatment regimens irrespective of the actual adherence, whereas the PP analysis assessed the effect of assigned treatment under full adherence. The results of the PP analysis suggest that participants who would have fully adhered to their assigned intensive blood pressure lowering therapy are expected to experience slightly more AKI recurrences than if they had fully adhered to the standard therapy in the first four years of the SPRINT. The ITT and PP effect estimates are very close to each other, which is consistent with our finding that full adherence to prescribed blood pressure medications, compared with observed adherence, does not substantially affect the recurrent outcome of interest in either the intensive or standard therapy arm.

%total effect ->  masked effect -> seperable effects 
We have accounted for the semi-competing risk of death on the recurrent outcome of AKIs while applying our methodologies on the SPRINT dataset. We estimated the total effect of treatment, which includes both the direct effect of treatment and the indirect effect mediated by death. {The estimated total effect closely mirrors the controlled direct effect, which could be attributed to the very low mortality rate of the study population ($<5\%$). However, controlled direct effects need to be interpreted with caution, because an estimand that eliminates death may not be well-defined and therefore the consistency assumption required for identification is likely violated \cite{young_compete_2020}.}

Given that there are marginally more expected AKIs in the intensive arm compared than the standard arm, this marginal difference may be due to more deaths in the standard arm than in the intensive arm \cite{wright2016randomized} and death prevents the occurrence of later AKIs.

Investigating the total effects remains meaningful as it reflects the real-world holistic impact of intensive versus standard blood pressure-lowering therapies on patients' AKI recurrences. 
The total effect reflects the overall effect of blood pressure lowering medications on AKI hospitalization, accounting for patients who die before developing AKI, and thus providing a measure of the treatments' real-world public health implications.

A potential limitation of the data analysis that may lead to biased estimation and/or incorrect inference is the definition of adherence. We assumed that the level of adherence of the participants at the beginning of each coarsened time interval reflects the true adherence of the participants to blood pressure medication during that interval.

This assumption is likely reasonable for SPRINT as participants' compliance rate is high in the study. However, in more general settings, if a participant does not fully adhere to the medications midway through a coarsened interval but is recorded as fully adhered in that interval, the coarsened adherence data would be inaccurate. Moreover, failing to consider possible adherence fluctuations within each time interval may open up new backdoor pathways given past covariates. 

\bibliographystyle{unsrt}
\bibliography{reference}

\newpage
\section*{Tables}

\begin{table}[H]
\centering
\caption{Summary of Key Notation}
\begin{tabular}{>{\raggedright\arraybackslash}p{2.5cm} p{12cm}}  % Adjust widths as needed
\toprule
\textbf{Symbol} & \textbf{Description} \\
\midrule
$N$ & Number of study subjects \\
$k$ & Time interval index, $k \in \{0, \dots, K\}$ \\
$L_0^*$ & A vector of baseline covariates measured at $k = 0$ \\
{$Z$} & {Baseline randomized treatment assignment, with $Z=1$ for treatment and $Z=0$ for control} \\
{$A_k, \ k>1$} & {Adherence indicator during interval $k$ relative to the assigned treatment $Z$, $A_k \in \{0,1\}$} \\
$L_{k}^*,\ k>1$ & A vector of time-varying covariates measured at interval $k$ \\
$L_{k}$ & An individual's joint time-varying covariates in interval $k$, composed of outcome and covariates: $L_{k} = (Y_{k}, L^*_{k})$\\
$Y_{k+1}$ & Observed binary outcome variable indicating event occurrence during interval $k+1$, $Y_{k+1} \in \{0,1\}$. \\
$R_{K+1}$ & Cumulative count of events over the study period: $R_{K+1} = \sum_{k=1}^{K+1} Y_k$  \\
{$\psi_{k}^{z,\bar{a}_{k-1}}$} & {Interval-specific counterfactual risk under assignment $z$ and adherence regime $\bar{a}_{k-1}$ through interval $k$: $\mathbb E(Y_k^{z,\bar{a}_{k-1}})$} \\
$P$ & True population distribution of the observed data. \\
\bottomrule
\end{tabular}
\label{tab:sec2_notation}
\end{table}

\vspace{2em}

\begin{table}[!htb]
\centering
\resizebox{\textwidth}{!}{
\begin{tabular}{cccc}
  \hline
    Time-varying Variables & Intensive Therapy &  Standard Therapy  & Overall \\   \hline
 Death Rate at $t=1$ ($D_1$) & 1.7\% & 2.0\% & 1.9\% \\ 
   Event Rate at $t=1$ ($Y_1$) & 2.7\% & 1.5\% & 2.1\% \\ 
  Average Mean Arterial Pressure at $t=1$ ($L_1$) & 85.0 & 95.0 & 90.0 \\   \hline
  Death Rate at $t=2$ ($D_2$) & 2.4\% & 3.2\% & 2.9\% \\  
  Event Rate at $t=2$ ($Y_2$)& 2.0\% & 1.3\% & 1.7\% \\ 
  Total Rate of Events ($R_2$) & 4.7\% & 2.8\% & 3.8\% \\
   \hline
\end{tabular}
}
\caption{Distributions of Key Time-varying Variables in the First Interval ($t=1$) and in the Second Interval ($t=2$) for the SPRINT Study Population Included in the Analysis ($n=9322$)}
\label{tb:tmle2_tv_sprint2}
\end{table}

\vspace{2em}

\begin{table}[!htb]
\begin{tabular}{cccc}
\hline
Section                 & Effect Estimand &    Parametric GLMs   &  XGB                   \\ \hline

\ref{sec:sprint_te} & ACE, ITT, TE  & 0.019 [0.011,0.026]  & 0.018 [0.010,0.027]  \\ \hline
\ref{sec:sprint_te}     & ACE, PP, TE     & 0.017 [0.007,0.027]  & 0.018 [0.009,0.028]  \\ \hline
\ref{sec:app_tmle2_apply_pie} & PIE, TE, $Z =1$  & 0.000 [-0.004,0.004] & 0.000 [-0.004,0.004]  \\ \hline
\ref{sec:app_tmle2_apply_pie} & PIE, TE, $Z =0$  & 0.003 [-0.001,0.006] & 0.001 [-0.002,0.004]  \\ \hline
\end{tabular}
\caption{Table of Estimated Causal Effects with 95\% Confidence Intervals Incorporating Parametric Generalized Linear Models (GLMs) and Extreme Gradient Boosting (XGB) Algorithms, repectively. {ITT: intention-to-treat;
PP: per-protocol; 
ACE: average causal effect; 
PIE: population intervention effect; 
CDE: controlled direct effect; 
TE: total effect;
{$Z$: initial assignment to intensive ($Z=1$) or standard ($Z=0$) blood pressure lowering target}
}}
\label{table:sprint_est_compare}
\end{table}

\end{document}